\documentclass[twocolumn,pre,aps,preprintnumbers,10pt,floatfix,showkeys]{revtex4-2}
\usepackage[utf8]{inputenc}
\usepackage[colorlinks=true, allcolors=blue]{hyperref}

\usepackage{amsmath}
\usepackage{amsfonts}
\usepackage{graphicx}
\usepackage[colorlinks=true, allcolors=blue]{hyperref}
\usepackage{tikz}
\usepackage{subcaption}

\newlength\imagewidth
\newlength\imagescale
\newcommand{\E}[2]{\mathbb{E}\left[{#2}\right]}

\newcommand{\var}[1]{\text{var}[{#1}]}

\newcommand{\kbt}[0]{k_{\text{B}}T}

\begin{document}
\title{Optimal Computation from Fluctuation Responses}
%\title{Locating the optimal computation protocol from fluctuation response relations}

\author{Jinghao Lyu}
\email{jolyu@ucdavis.edu}
\affiliation{Complexity Sciences Center and Physics and Astronomy Department,
University of California at Davis, One Shields Avenue, Davis, CA 95616}

\author{Kyle J. Ray}
\email{kjray@ucdavis.edu}
\affiliation{Complexity Sciences Center and Physics and Astronomy Department,
University of California at Davis, One Shields Avenue, Davis, CA 95616}

\author{James P. Crutchfield}
\email{chaos@ucdavis.edu}
\affiliation{Complexity Sciences Center and Physics and Astronomy Department,
University of California at Davis, One Shields Avenue, Davis, CA 95616}

\begin{abstract}
The energy cost of computation has emerged as a central challenge at the intersection of physics and computer science. Recent advances in statistical physics---particularly in stochastic thermodynamics---enable precise characterizations of work, heat, and entropy production in information-processing systems driven far from equilibrium by time-dependent control protocols. A key open question is then how to design protocols that minimize thermodynamic cost while ensuring correct outcomes. To this end, we develop a unified framework to identify optimal protocols using fluctuation response relations (FRR) and machine learning. Unlike previous approaches that optimize either distributions or protocols separately, our method unifies both using FRR-derived gradients. Moreover, our method is based primarily on iteratively learning from sampled noisy trajectories, which is generally much easier than solving for the optimal protocol directly from a set of governing equations.  We apply the framework to canonical examples---bit erasure in a double-well potential and translating harmonic traps---demonstrating how to construct loss functions that trade-off energy cost against task error. The framework extends trivially to underdamped systems, and we show this by optimizing a bit-flip in an underdamped system. In all computations we test, the framework achieves the theoretically optimal protocol or achieves work costs comparable to relevant finite time bounds. In short, the results provide principled strategies for designing thermodynamically efficient protocols in physical information-processing systems. Applications range from quantum gates robust under noise to energy-efficient control of chemical and synthetic biological networks.
% \jlnote{Need more description on the flip protocol}
\end{abstract}

\date{\today}

\preprint{arxiv.org:2509.XXXXX [cond-mat.stat-mech]}


\maketitle

\section{Introduction}
Nowadays, computation—ranging from operating everyday digital devices to training large-scale machine learning models—consumes a significant amount of energy. This raises questions we long to answer: \textit{how to perform computation efficiently} and \textit{what is the minimal cost of implementing computation?} The most well-known result in this setting is Landauer's bound that the minimum energetic cost to erase one bit of information at temperature $T$ is $\kbt \log2$, where $k_{\text{B}}$ is Boltzmann's constant \cite{landauer1961irreversibility}. Though Landauer’s original derivation has limitations, it was groundbreaking in establishing a deep connection between computation and thermodynamics.

Since Landauer's time, we now appreciate that rigorously investigating the energetic costs of computation requires nonequilibrium physics---as all computation takes place in physical systems operating far from equilibrium. As a key innovation in this, researchers successfully developed a stochastic dynamics for thermodynamic resources. This, in turn, led to \textit{stochastic thermodynamics} \cite{seifert2012stochastic} and established an arena for exploring the thermodynamics of computation.

Consider a stochastic system $\mathcal{S}$ with a time-dependent energy landscape control parameter $a_{t}$. Suppose the initial distribution over  $\mathcal{S}$ is $p_{0}$. To implement a computation we aim to transform the distribution from $p_{0}$ to a goal distribution $p_{\text{target}}$ over a time duration $\tau$. We further impose a cyclic boundary condition on the control protocol---$a_{0} = a_{\tau}$---to enable information storage and to support follow-on computations. The question then becomes: how to design the time-dependent protocol $a_t$ to transform the distribution from $p_{0}\to p_{\text{target}}$ at minimal energetic cost.

Two distinct energetic cost optimization approaches have been studied previously \cite{schmiedl2007optimal, gomez2008optimal, zulkowski2014optimal,gingrich2016near, dechant2019thermodynamic, zhong2022limited, proesmans2020finite, proesmans2020optimal,  boyd2022shortcuts, dago2023logical, engel2023optimal,whitelam2023demon,sabbagh2024wasserstein, davis2024active, sanders2024optimal, loos2024universal, klinger2025minimally, rengifo2025machine, sanders2025minimal, gupta2025thermodynamic}. The first is distribution-oriented: transform an initial distribution into a target distribution, without regard to the final energy landscape. In this, we have the correct distribution but the potential is not correct. As a result, the information after the processing is unstable and cannot be used in the future. The second approach is protocol-oriented: drive the system from one protocol configuration to another, without considering the resulting final distribution.  However, this is unsuited to computational tasks. Imposing constraints on both final distribution and final protocol values can over-constrain the system, leading to no feasible solution at all. 

%These two classes of boundary conditions, however, are \textit{not} compatible with the computation tasks. The first class of boundary conditions leads to the desire distribution indeed but the final energy landscape is uncontrollable. The computation tasks usually require a cyclic protocol: $a_{0}=a_T=a$. So we need to introduce sudden changes in the protocol values at the end of computation which incurs additional costs. The second setup only optimizes costs and will lead to a wrong distribution or results. Furthermore, the cyclic boundary conditions will enforce a constant control protocol $a_{t}=a$ since the cost with a constant protocol is simply $0$. So far, we still do not know how to optimize the protocol in computational tasks. 

Unlike previous approaches that only optimize either distributions or protocols, we introduce a framework integrating both optimizations. Following Ref. \cite{deffner2013information} we use a coarse-grained distribution to store information. In a computational task, we do not ask the final distribution in $\mathcal{S}$ to be exactly $p_{\text{target}}$. Instead, we only need the coarse-grained distribution at final time ${p_{\tau}}_{\text{cg}}$ to match the coarse-grained target distribution ${p_{\text{target}}}_{\text{cg}}$. Therefore, we can ensure both a correct coarse-grained distribution and a correct potential landscape. We show how to design loss functions for different computation tasks and use fluctuation response relations (FRR) to compute the loss function gradients. We demonstrate that these loss functions successfully drive the distribution toward the target coarse-grained distribution, all the while achieving work costs on the order of Landauer's lower bound.
%In this work, we propose a novel framework to find the optimal control protocol of computational tasks by mixing up these two boundary conditions. The foundation of our framework is fluctuation-response relations  For each computational task, we design a current as a loss function. Instead of using thermodynamic cost alone as loss functions, we add a key term---logical error loss---in our loss function. Then we use the gradient from fluctuation-response relations to find the optimal solution. 

Our framework has several advantages: (1) Based on universal fluctuation-response relations, it can be used to optimize a large family of stochastic systems including overdamped and underdamped Langevin dynamics, Markovian jump processes, and the like. (2) It allows for highly flexible weighting of various target metrics, like logical error and energetic cost, in the loss function. (3) Because the learning algorithm works by gradient descent and observing samples of simulated trajectories, rather than solving a system of equations, it lends itself to computer implementation that takes advantage of modern machine learning tools that can scale to higher dimensional problems.
% \jlnote{Maybe more?}

The development is organized as follows: Sec. \ref{sec:FRrelation} reviews FRR basics and turns to discuss loss-function design in Sec. \ref{sec:Lossdesign}. Then we use FRR to optimize loss functions and explore numerical examples. Throughout, $(\cdot)_t$ denotes a quantity value $(\cdot)$ at time $t$. 

\section{Fluctuation Response Relation}
\label{sec:FRrelation}

We first review the fluctuation-response relation (FRR) in the context of overdamped dynamics \cite{warren2012malliavin, zheng2024universal, klinger2025computing}. Consider one-dimensional overdamped Langevin dynamics with an external control parameter $a$:
\begin{align}
    dx &= \mu\Phi(x,a) dt + \sigma dw_t~,
\end{align}
where $\mu$ is the particle mobility, $dw_t$ is an infinitesimal Wiener process, $\sigma=\sqrt{2\mu\kbt}$ is the standard deviation of the thermal noise or $D = \frac{1}{2}\sigma^2$ is the diffusion constant, and $\Phi(x,a)=-\partial_{x}U(x,a)$ with $U(x,a)$ being the potential of the system as a function of the position $x$ and the control parameter $a$.

An observable (called a cumulant current) $J$ along a trajectory $\omega=\{x_{t}\}_{t=0}^{\tau}$ is defined as:
\begin{align}
    J(\omega,a)=\int _{\omega} f(x,a) \cdot dx + g(x,a) dt~,
\end{align}
where $\cdot$ is the It\^{o} product. For example, the energetic cost (work) of one trajectory $\omega$ can be written as $W(\omega,a) = \int_{\omega} \partial_{t}U(x,a) dt$. The expectation value of $J(\omega,a)$ over an ensemble of trajectories is:
\begin{align}
    \E{\omega}{J(\omega,a)} = \sum_{ \omega}\text{Pr}(\omega,a) J(\omega,a)~,
\end{align}
where $\text{Pr}(\omega,a)$ is the probability of observing the trajectory $\omega$ under protocol $a$. 

The purpose of an FRR is to monitor how the expectation of a current $J$ changes due to variation in the control parameters: $\frac{\partial }{\partial a}\E{\omega}{J(\omega,a)}$. From the chain rule, we have:
\begin{align}\label{eqn:FRR1}
&\partial_{a}\E{\omega}{J(\omega,a)}=\E{\omega}{\partial_{a}J(\omega,a)}+\E{\omega}{J(\omega,a)\partial_{a}\log \text{Pr}(\omega,a)}
  ~.
\end{align}
The first term can be readily calculated given a particular form of $J$ and a specific protocol. The second term, involving derivatives of the score function, requires special attention. We can compute this term from trajectory data by noting that the probability of observing a trajectory $\omega$, using the the Onsager-Machlup action formalism, can be written as \cite{onsager1953fluctutations}:
\begin{align}\label{eqn:od_omegaprob}
    \text{Pr}(\omega,a) = \mathcal{N} \exp\left\{-\frac{1}{2\sigma^2}\int_{\omega}[\dot{x}-\mu\Phi(x,a)]^2dt\right\}~,
\end{align}
where $\mathcal{N}$ is a trajectory independent constant that plays no role in $\partial_a \log \Pr(\omega,a)$:
\begin{align}\label{eqn:od_logomegaprob}
    \partial_a\log \text{Pr}(\omega,a) =\frac{1}{\sigma^2}\int_{\omega} \mu\partial_a\Phi(x,a)[\dot{x}-\mu\Phi(x,a)] dt
    ~.
\end{align}
From the Langevin equation, we recognize that $[\dot{x}-\mu\Phi(x,a)] dt = \sigma dw_t$. Thus, plugging the above into Eq. (\ref{eqn:FRR1}) yields:
\begin{align} \label{eqn:FRR}
&\partial_{a}\E{\omega}{J(\omega,a)} = \nonumber \\
&~~\E{\omega}{\partial_{a}J(\omega,a)}+\E{\omega}{J(\omega,a)\frac{1}{\sigma}\int_{\omega}\mu\partial_{a}\Phi(x,a)\cdot dw_t}.
\end{align}
%The expectation with protocol $a'_{t}=a_{t}+\deltaa_{t}$ is 
%\begin{align}
%    \E{J}_{a'}& = \sum_{\omega}\text{Pr}_{a}(\omega) J(\omega,a') \frac{\text{Pr}_{a'}(\omega)}{\text{Pr}_{a}(\omega)}\nonumber\\
%    &=\E{J(\omega,a') \frac{\text{Pr}_{a'}(\omega)}{\text{Pr}_{a}(\omega)}}_a~.
%\end{align}
%We see that the perturbed expectation is reformulated as the expectation value in the original dynamics with protocol $a$. Next, we express the $J(\omega,a') \frac{\text{Pr}_{a'}(\omega)}{\text{Pr}_{a}(\omega)}$ in terms of the original protocol $a$. For the current $J(\omega,a')$, it is simple
%\begin{align}
%    J(\omega,a') =J(\omega,a) + \frac{\partial}{\partial a}J(\omega ,a)\delta a + \dots
%\end{align}
%The probability ratio can be written as 
%\begin{align}
%    \frac{\prpert{\omega}}{\prog{\omega}}=\exp\left[\int_{\omega} \frac{\delta \Phi}{\sigma}dw+\frac{1}{2}\int_{\omega}\frac{\delta\Phi^2}{\sigma^2}\right]
%\end{align}
%where $\delta \Phi= \Phi(x,a'_t)-\Phi(x,a_t)$.
%Up to the first order of $\delta a$, we have
%\begin{align}
%    \partial_{a}\E{J} = \E{\partial_{a} J} +\E{{\sigma}^{-1}J\int_{\omega}\partial_{a}\Phi \cdot dw}~.
%\end{align}
%The integral $\sigma^{-1}\int_{\omega} \partial_{a}\Phi dw$ is called the Malliavin weight. 

This is the FRR, which relates the thermal noise in the observed trajectories to the response function. It can also be derived from the Girsanov theorem \cite{liptser2013statistics}. The FRR allows us to easily access the gradient of observables with respect to control parameters through trajectory simulation.

\section{Loss functions in Computations}\label{sec:Lossdesign}

Having introduced the essentials behind the FRR, we now associate the observable $J$ with a loss function designed to be minimal at some desired behavior. From this perspective, the FRR is a flexible and powerful tool to apply machine learning methods to noisy physical systems: we can use it to learn control protocols that push a system of interest to behave in targeted ways. We now discuss how to design a loss function specifically useful for computational tasks. To begin with, we must establish how information is encoded within a physical system.

Consider a classical continuous $n-$dimensional system state $\mathcal{S} \in \mathbb{R}^{n}$. Encoding binary information requires coarse-graining $\mathcal{S}$---partitioning its continuous space into distinct regions that represent discrete-value states. Mathematically, a coarse-graining ${\sigma}$ is a many-to-one map from a distribution $p$ over system $\mathcal{S}$ to a distribution $p_{\mathcal{I}}$ over an informational space $\mathcal{I}$. The informational space $\mathcal{I}$ is a finite dimension space spanned by binary values; i.e., $\text{b}=\{|00000\rangle,|00001\rangle, \dots\}$. After coarse-graining, the degrees of freedom over this finite dimensional space are called \textit{information-bearing degrees of freedom} or IBDs \cite{deffner2013information}.

To have a stable information register for $n-$binary digits, we ask the original potential to have $2^{n}$ minima $\{x_{\text{b}}\}$, each of which represents $n$ binary digits. A simple example is a double-well potential $V(x) = \frac{1}{4}x^4-\frac{1}{2}x^2$ with two stable minima located at $x^{*}=\pm1$. Each minimum can be used to store a binary digit. We use the phase space regions surrounding $x=-1$ and $x=+1$ to represent $|0\rangle$ and $|1\rangle$, respectively. The coarse-graining map sends any distribution $f(x)$ over $\mathbb{R}$ to the distribution $p_{\mathcal{I}} = (p_{|0\rangle}, p_{|1\rangle})$, where $p_{|0\rangle} = \int_{-\infty}^{0} p(x)dx$ and $p_{|1\rangle} = \int_{0}^{+\infty} p(x)dx$.

Despite the eventual goal of digital computation, we will not derive effective dynamics describing the distribution evolution over IBD but rather optimize the dynamics of the continuous underlying physical system $\mathcal{S}$. The aim is to design protocols such that the system's final distribution $f(x_\tau)$ maps to the correct distribution ${p_{\mathcal{I}}}_{\text{target}}$ with minimal energetic cost. To do this, we prioritize achieving the target distribution over IBD, minimizing the work cost simultaneously. This motivates the loss function:
\begin{align}\label{eqn: total_loss}
    \mathcal{L}_{\text{loss}}(a) = \E{\omega}{\alpha_{1}\mathcal{L}_\text{error}(\omega,a)+ \alpha_W{W(\omega,a)}}~,
\end{align}
where $\mathcal{L}_{\text{error}}$ represents an appropriate error measurement in $\mathcal{S}$ and where $\alpha_{1}$ and $\alpha_{W}$ are hyperparameters. %The role of $W(\omega,a)$ is analogous to the ridge regression in linear regression, where $L_{2}$-norm is used to balance fitting the data and the model complexity. 

Suppose we wish to perform a computation task sending bit value $\text{b}_{i}$ to $g(\text{b}_{i})$. We ask that in the physical system $S$, all trajectories starting around the minimum $x_{\text{b}_{i}}$ representing $\text{b}_{i}$ to end around the minimum $x_{g(\text{b}_i)}$ representing $g(\text{b})$. To enforce this behavior, we choose:
\begin{align}\label{eqn: error_loss}
    \mathcal{L}_{\text{error}} = (x_{\tau}-x_{g(\text{b}_i)})^2~, \text{ if } x_0\text{ near }x_{\text{b}_{i}}~.
\end{align}
We add higher-order moments $\E{\omega}{(x_{\tau}-x_{g(\text{b}_i)})^n}$ to the loss function if the L2 norm is insufficient.

Before diving into specific examples, we should discuss the optimization framework underlying this intuitively-motivated loss function. The straightforward approach to achieve energetically-efficient operations that meet an accuracy threshold is to set an upper bound on error rate $\epsilon$ and then locate the most energetically-efficient protocol with error being no greater than $\epsilon$. In optimization this is the \emph{primal problem} \cite{boyd2004convex}:
\begin{align*}
    \underset{a}{\text{minimize~}}\E{\omega}{W(\omega,a)}
    \text{~subject to ~} \E{\omega}{\mathcal{L}_\text{error}(\omega,a)}- \epsilon\leq0~.
\end{align*}
The corresponding Lagrangian \textit{dual problem} is: 
\begin{align*}
    &\underset{\lambda}{\text{maximize~}} \inf_a {\left(\E{\omega}{W(\omega,a)}+\lambda (\E{\omega}{\mathcal{L}_\text{error}(\omega,a)}-\epsilon)\right)} ~,
\end{align*}
subject to $\lambda \geq 0$.

Suppose the optimal value of the primal problem (the optimal work cost with error being no greater than $\epsilon$) is $p^*$ and that of the Lagrangian dual problem is $d^*$. If strong duality holds, these two problems are equivalent to each other $p^* = d^*$, i.e., for any $\epsilon$ we can find a $\lambda$ such that these two optimizations are equivalent. If we do not have strong duality, then we can simply pick a $\lambda$ to optimize over and get a lower bound on the primal problem energetic cost $d^* \leq p^*$. 

% \section{Examples}\label{sec:Example} 

\paragraph*{Example: Overdamped Harmonic Trap Translation}
\label{sec:Example} 

We start with a well-known model for which the optimal protocol can be solved exactly: moving a harmonic trap from $0$ to $a_f$ over the time duration $\tau$. The potential is $V(x,t)=(x-a_t)^2/2$, where \(a_t\) is the center of the harmonic well and we have full control of $a_{t}$ with boundaries fixed at $a_0=0$ and $a_\tau=a_f$. The corresponding stochastic differential equation of a unit-mobility particle is:
\begin{align}\label{eqn:move_sde}
    dx= -(x-a_t)dt+\sigma dw_t~.
\end{align}
We assume the system is in equilibrium at $t=0$.

In addition to minimizing the work, the particle must end up as close as possible to the final target position $a_f$. The loss function is taken to be:
\begin{align}
    \mathcal{L}_{\text{move}}(a_{t}) = \alpha_1\E{\omega}{(x_\tau-a_f)^2} + \alpha_{W}\E{\omega}{W(\omega,a_{t})}~.
\end{align}
Our goal now is to find the control protocol $a^{*}_{t}$ that minimizes the loss function $\mathcal{L}_{\text{move}}$. This can be done analytically using the method of Lagrange multipliers. (See the Appendix \ref{appendix: translation_details}.) In short, the optimal protocol is that $a_{t}$ is a linear function of time $t$ except at times $t=0$ and $t = \tau$.
%Let us work out several examples. The first study the sudden jump protocol:
%\begin{align}
%    a(t) = a_{f}a(t-1)~.
%\end{align}
%The work can be computed from the average of  $\partial_{t}V = -\partial_{t} a (x-a(t)) = \delta(t-1)a_{f}(x-a(t))$. Since the mean of the position $x$ cannot jump. The work is $a_{f}^2$. 
%Next, we study a linear protocol:
%\begin{align}
%    a(t) = a_{f} t~.
%\end{align}
%The final position mean and the work mean can be written in closed form:
%\begin{align}
%    \E{\omega}{x_\tau}=\int_{0}^{\tau}e^{-(\tau-t)}a_t dt~,~ \E{\omega}{W}=\int_{0}^{\tau} (a_t-\mu_t)\dot{a}_t dt~ . 
%\end{align}
%with constraints $\dot{\mu}_t + \mu_t =a_t$ from averaging Langevin equation Eq. \eqref{eqn:move_sde}. The Euler Lagrange equation gives $\ddot{\mu}_t= 0$.
%the Euler-Lagrange equations yield a family of optimal solutions:
%\begin{align}
%    \E{\omega}{x_\tau}= mt~, ~a_t=mt+m~.
%\end{align}

\begin{figure}
    \centering
    \includegraphics[width=0.9\linewidth]{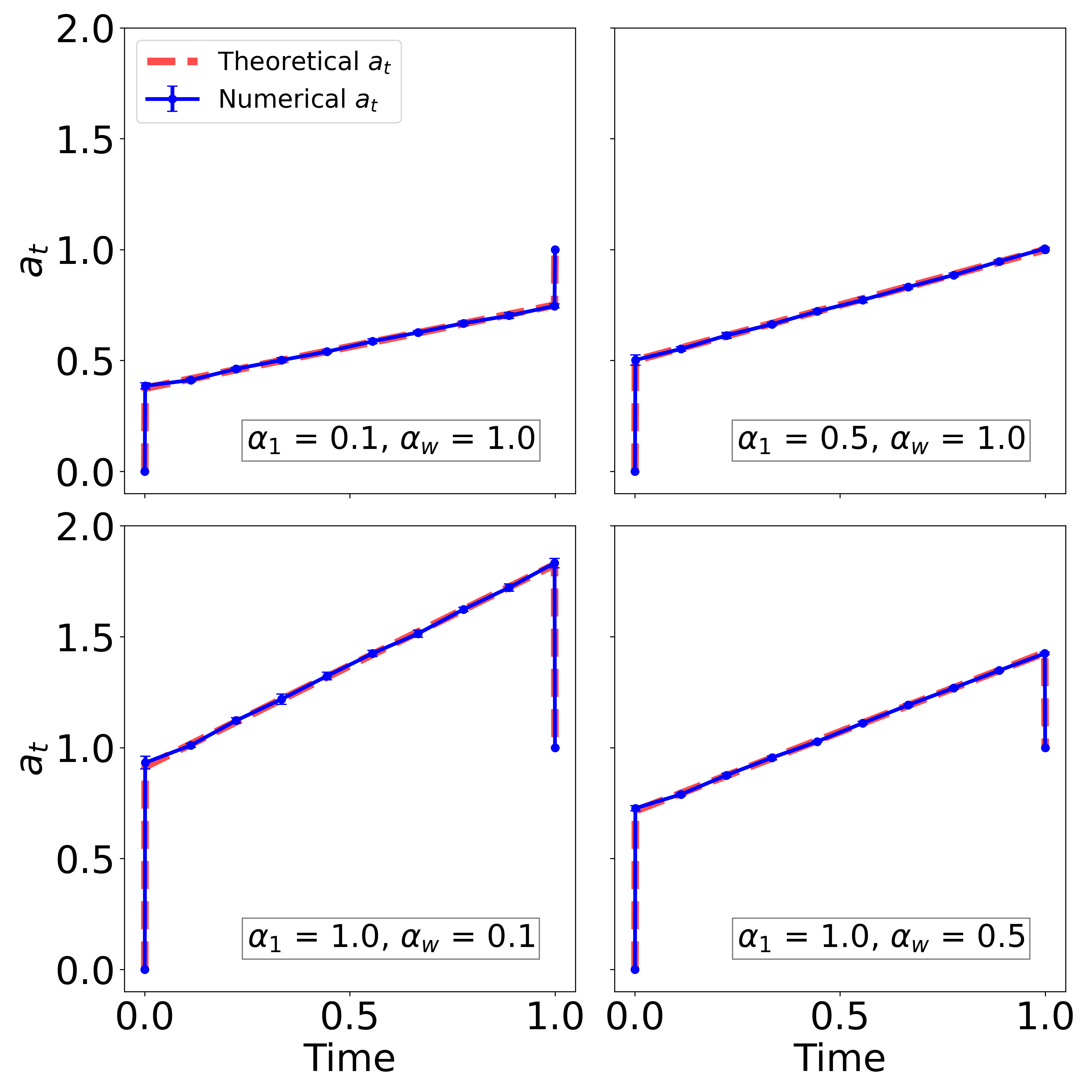}
\caption{Optimal protocol for translating a harmonic trap: We fix $a_f=1$ and $\tau=1$ and use $10$ break points. For one training step, we simulate $5000$ trajectories with protocol value $\boldsymbol{a}=\{a_{1}, \dots, a_{10}\}$ to compute the gradient of the loss functions and use the gradient to update $\boldsymbol{a}$ and we train the protocol for $600$ iterations. We numerically show different types of jump with different weight combinations.}
    \label{fig:move_results}
\end{figure}

For numerical optimization, we parameterize the protocol values $a_{t}$ by a continuous piecewise-linear function with parameters that are the break points $\boldsymbol{a}=\{a_{1},\dots,a_{n}\}$. We calculate the gradient $\partial_{\boldsymbol{a}} \mathcal{L}_{\text{move}}(\boldsymbol{a})$ with simulated trajectories in each training iteration and update $\boldsymbol{a}$ for the next iteration. Figure \ref{fig:move_results} shows the training results with different weights in the loss function.

\paragraph*{Example: Bit Erasure}

\begin{figure}
    \centering
    \includegraphics[width=1\linewidth]{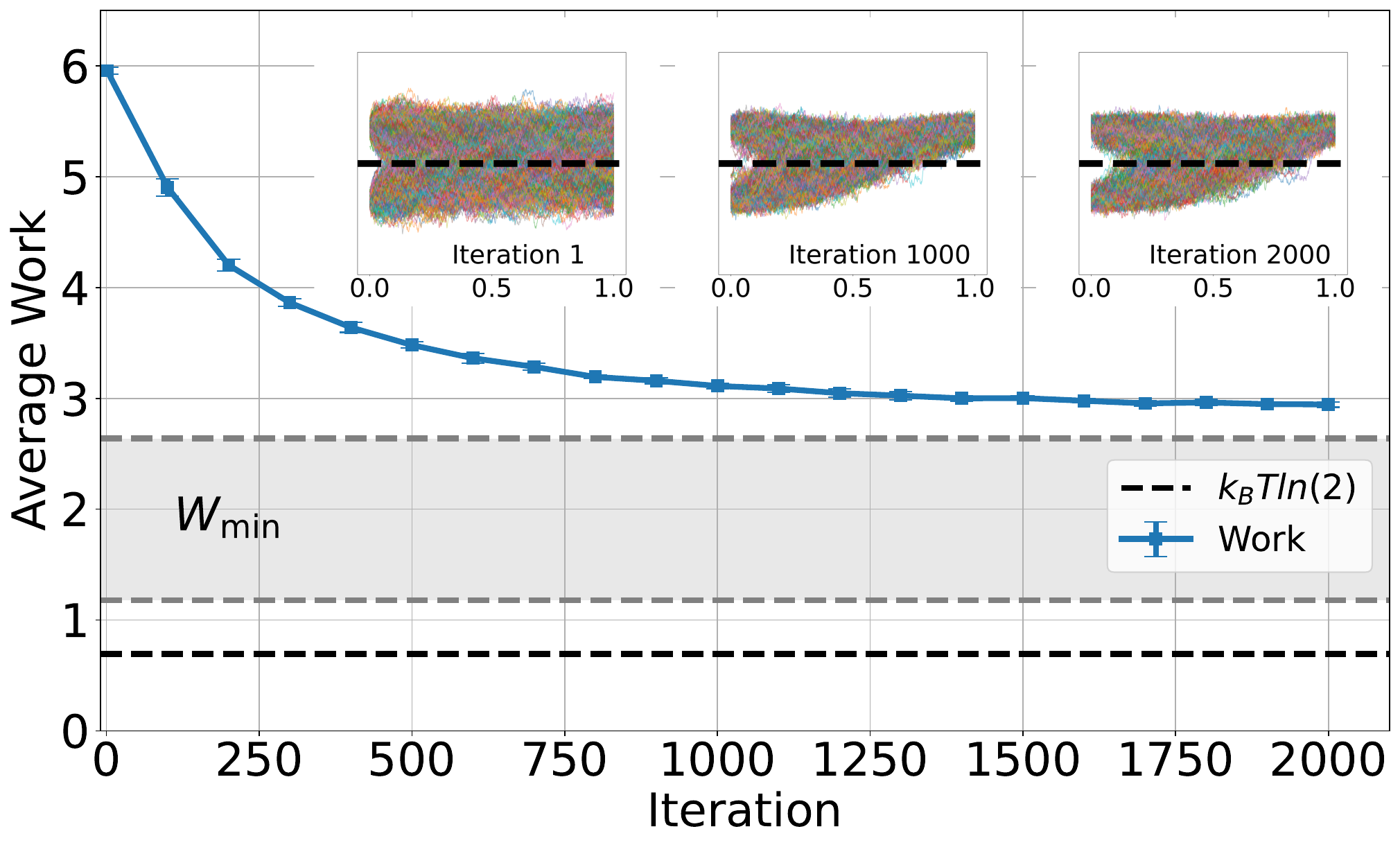}
\caption{Bit erasure optimization: For each training iteration, we simulate $5000$ trajectories to compute the gradient and update parameters. We validate with $10^5$ trajectories to estimate work cost and error. The number of error trajectories almost vanishes after $100$ iterations. Three trajectory ensembles are selected from a random training. The plot also includes the upper and lower bound of the minimal work cost in finite-time bit erasure under full control of the potential \cite{proesmans2020finite}, denoted with a gray highlight.}
    \label{fig:erasure_resultsI}
\end{figure}

\begin{figure}
    \centering
    \begin{subfigure}{1\columnwidth}
        \subcaption{}\includegraphics[width=1\linewidth]{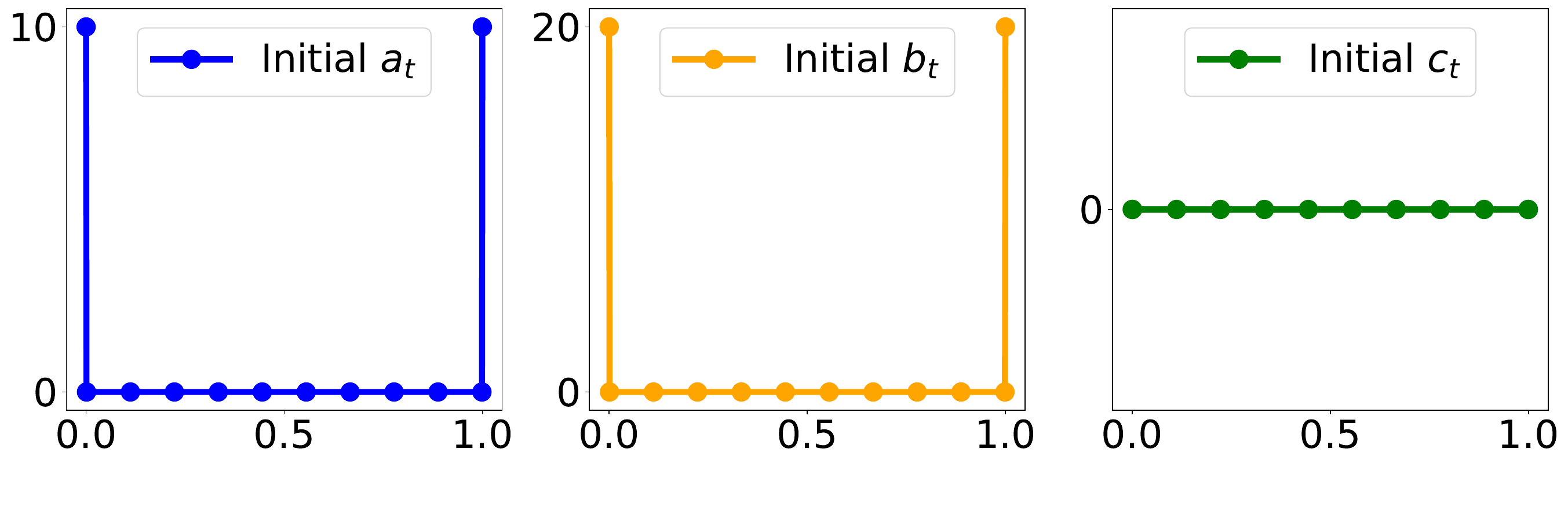}
    \end{subfigure}
    \begin{subfigure}{1\columnwidth}
        \subcaption{}\includegraphics[width=1\linewidth]{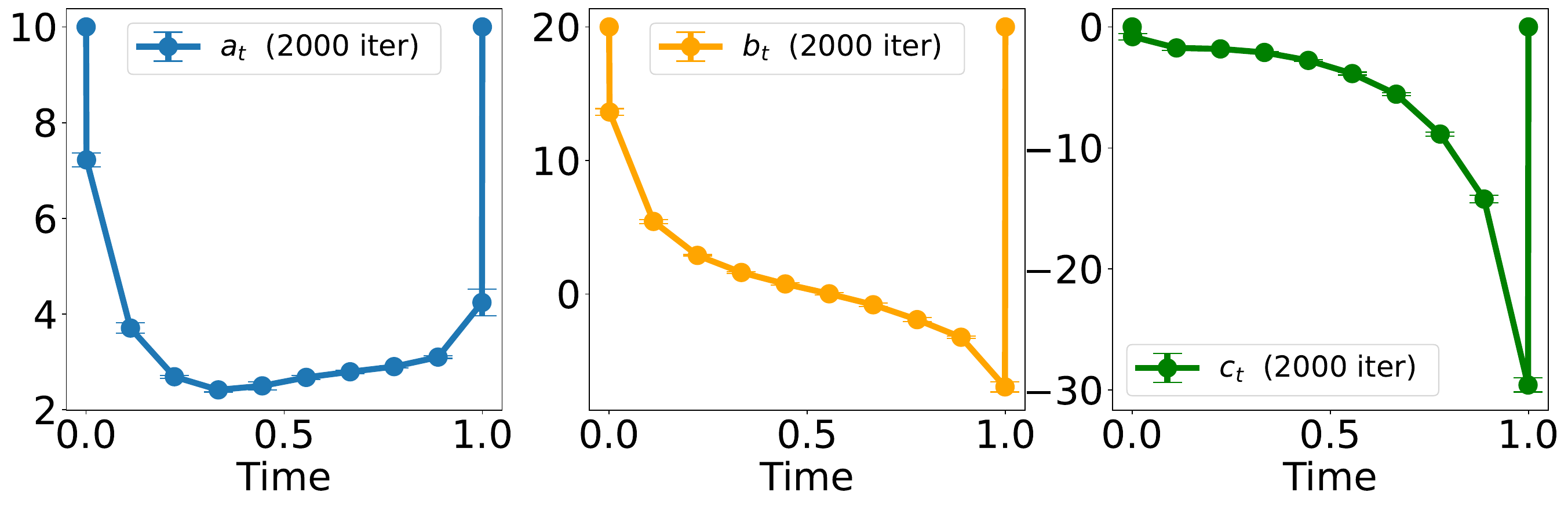}
    \end{subfigure}
    \begin{subfigure}{1\columnwidth}
        \subcaption{}\includegraphics[width=1\linewidth]{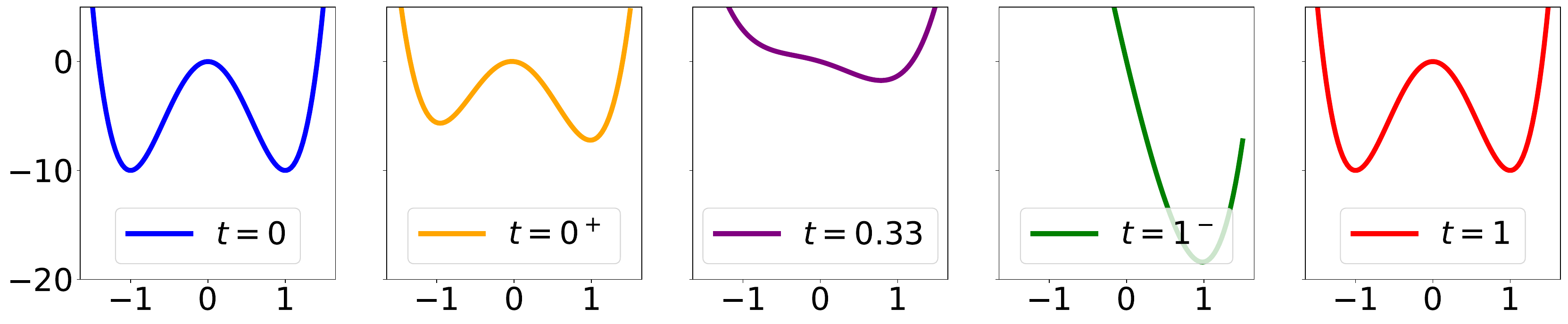}
    \end{subfigure}
\caption{Optimal erasure protocol: For each protocol parameter \{$a_{t}$, $b_t$, $c_t$\}, we use $10$ break points for parameterization. (a) They are initialized to $0$. (b) Final protocol values after training for $1000$ iterations. (c) Potentials $U$ at different times.}
    \label{fig:erasure_resultsII}
\end{figure}

We now move into the domain of optimizing computations by considering single-bit erasure. Bit erasure is a fundamental computational task that maps the uniform distribution over IBD ${p_\mathcal{I}}_0 = (1/2,1/2)$ to a nearly-deterministic state ${p_\mathcal{I}}_\text{target} = (\epsilon,1-\epsilon)$. To implement bit erasure, we use a quartic potential with three parameters $\boldsymbol{a}_t=\{a_{t}, b_{t}, c_{t}\}$:
\begin{align}
    U(x,t) = a_t x^4 - b_t x^2+ c_t x~.
\end{align}
The parameter boundary conditions are cyclic, with $a_0=a_{\tau}=a,~b_0=b_{\tau}=b,~c_{0}=c_{\tau}=0, \text{ and } 2a=b$. Parameters $a_t$ and $b_t$ control equilibrium positions $x_{\text{eq}}=\pm b_{t}/2a_t$ and $c_t$ tilts the potential. With these boundary conditions imposed, the potential $U$ can encode a binary value as it has two stable equilibrium points $x=\pm 1$. The particle is at equilibrium with potential $U(x,0)$ at temperature $T$ initially. At time $t=\tau$, all particles are required to be around $x=+1$, so we choose the loss function in bit erasure task to be:
\begin{align}
    \mathcal{L}_{\text{erase}}(\boldsymbol{a}_t) =  \E{\omega}{\alpha_{1}\left(x_\tau-1\right)^2+\alpha_{W} W(\omega, \boldsymbol{a}_t)}~.
\end{align}
In numerical experiments, we set $\alpha_{1}=1$, work weight $\alpha_W=0.1$, $\mu=1$, $\kbt=1$, and $\tau=1$. Figures \ref{fig:erasure_resultsI} and \ref{fig:erasure_resultsII} show the numerical optimization results. 

In the quasistatic limit, efficient erasure operates as follows. First, lower the energy barrier; next, tilt the potential; finally, raise the energy barrier and remove the tilt. In our optimized finite-time protocol, we see that lowering the barrier and tilting the potential happen all at once at $t=0^{+}$. Similar jumps in optimal protocol are also found in Refs. \cite{proesmans2020finite, proesmans2020optimal}. The linear potential $c_{t}x$, used to drive the particle towards $x=+1$, dominates after $t=0.33$ and $U$ is almost linear in the region of interests $(x\in[-1,1])$. After $t=0.33$, $c_{t}$ decreases quickly to around $-30$. $c_{1^{-}}\simeq -30$ ensures that the  potential $U(x,t=1^{-})$ has minima around $x=+1$, aligned with the minima at the fixed boundary conditions. The work cost after $2000$ iterations is $\Tilde{W}_{\min}=(2.942\pm0.026 )\kbt$.

To compare our result with other attempts at bit erasure, we compute the ratio $r=(\Tilde{W}_{\min} - \ln2)/(\var{x_{0}}/D\tau)=2.311\pm0.027$. The ratios of other bit erasure frameworks \cite{berut2012experimental, berut2013detailed, jun2014high, zulkowski2014optimal, berut2015information,gavrilov2017direct, boyd2022shortcuts} range from $2.89$ to $5.67$. (See the detailed table in Ref. \cite{proesmans2020finite}.) While we cannot analytically calculate the optimal protocol for this setting, we see that the training converges to a work cost that is close to the upper bound $r=2$ for optimal erasure under full control of the potential \cite{proesmans2020finite}. Given that our control is limited by both our choice of piecewise linear protocol and the form of the potential and the closeness of our distribution at $t=\tau$ to the local equilibrium distribution, this proximity to the work cost for unrestricted optimal control is evidence that our learning algorithm is finding truly efficient protocols.

We end this example by a brief discussion of the optimal protocol's dependence on the hyperparameter ratio $\alpha_{W}/\alpha_1$. Increasing the ratio $\alpha_{w}/\alpha_{1}$ can lower the work at the cost of higher error rates. To understand this, consider the following two reference protocols. In Protocol 1, the potential $U(x,t)$ remains unchanged during $[0,\tau]$ so that $\E{}{W}\sim 0$ but the error is $\E{}{\mathcal{L}_{\text{error}}}\sim 0^2\cdot\frac{1}{2}+2^2\cdot\frac{1}{2}=2$. Protocol 2 is a faithful erasure for which $\E{}{W}\sim \log2 + 2\var{x_{0}}/(D\tau)=2.69$ \cite{proesmans2020finite} and $\E{}{\mathcal{L}_{\text{error}}}\sim 0$. Around $\alpha_{w}/\alpha_{1} \sim 0.7$, the two protocols are equally favorable, with Protocol 2 being more favorable below this value. Numerically, starting from $\alpha_{w}/\alpha_{1} = 0.1$, we observe a sharp transition from Protocol 2 to Protocol 1 around $0.47$, strongly suggesting a second-order phase transition, which we leave for detailed future study.

\begin{figure}
    \centering
    \includegraphics[width=1\linewidth]{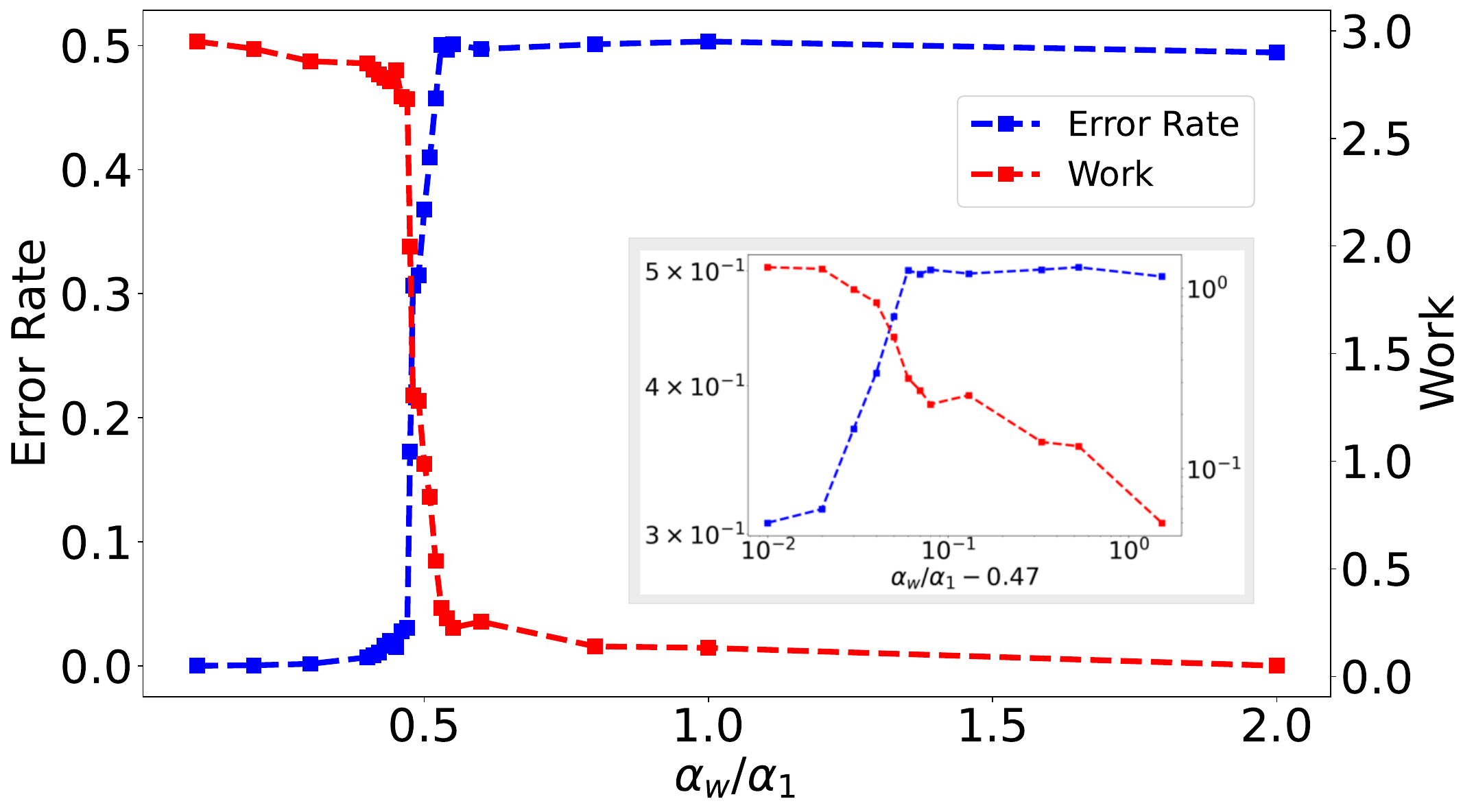}
    \caption{Error rate and work dependence on the hyperparameter ratio in erasure: Around $\alpha_{w}/\alpha_{1}= 0.47$, we observe a sharp rise in error rate. The inset is error rate and work versus $(\alpha_{w}/\alpha_{1}- 0.47)$ in log scale, indicating a second order phase transition.}
    \label{fig:erasure_phase}
\end{figure}

\begin{figure}
    \centering
    \begin{subfigure}[t]{0.99\columnwidth}  
        \subcaption{} \includegraphics[width=1.0\columnwidth]{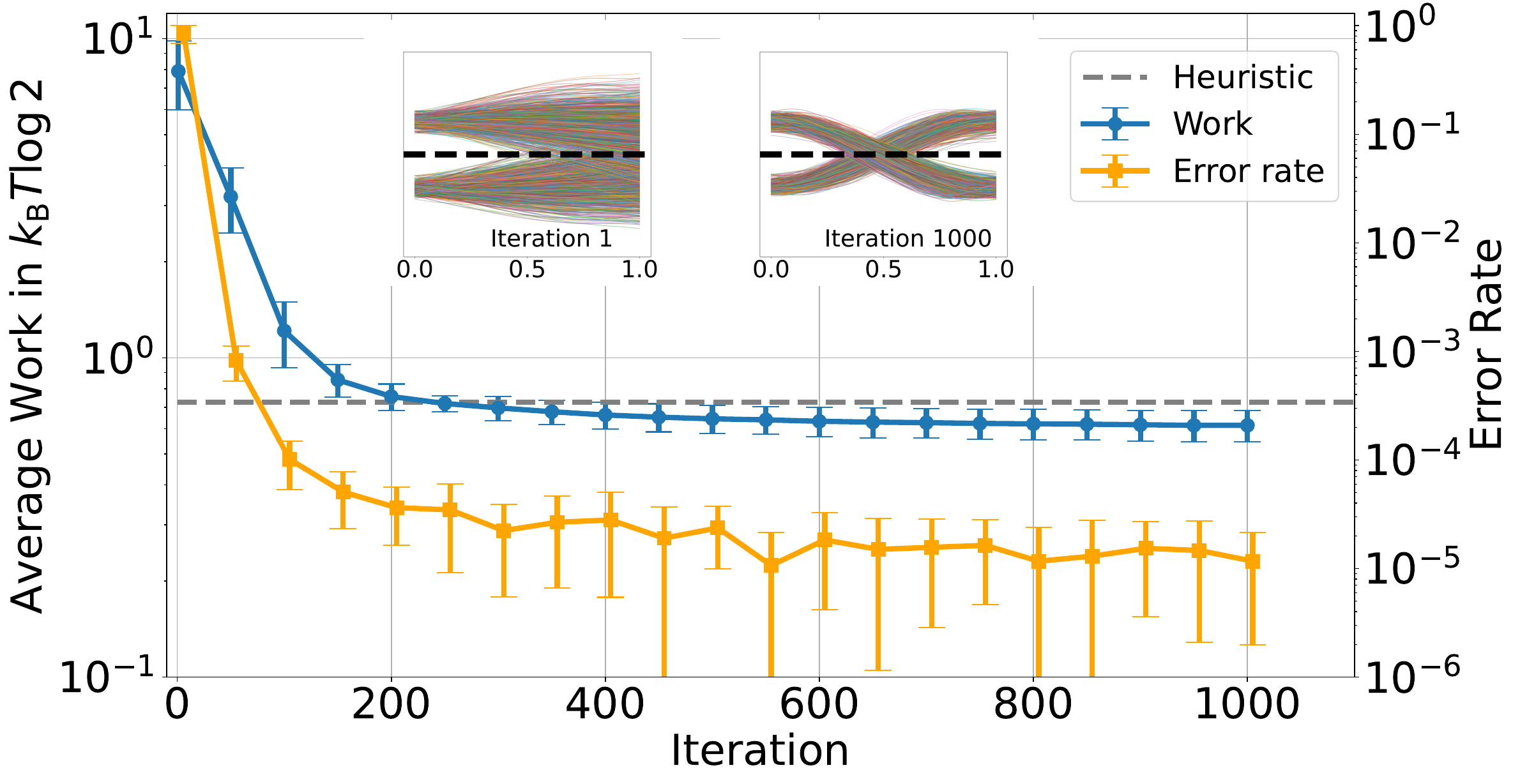}
    \end{subfigure}
    \begin{subfigure}[t]{1.0\columnwidth}
        \subcaption{} \includegraphics[width=1\columnwidth]{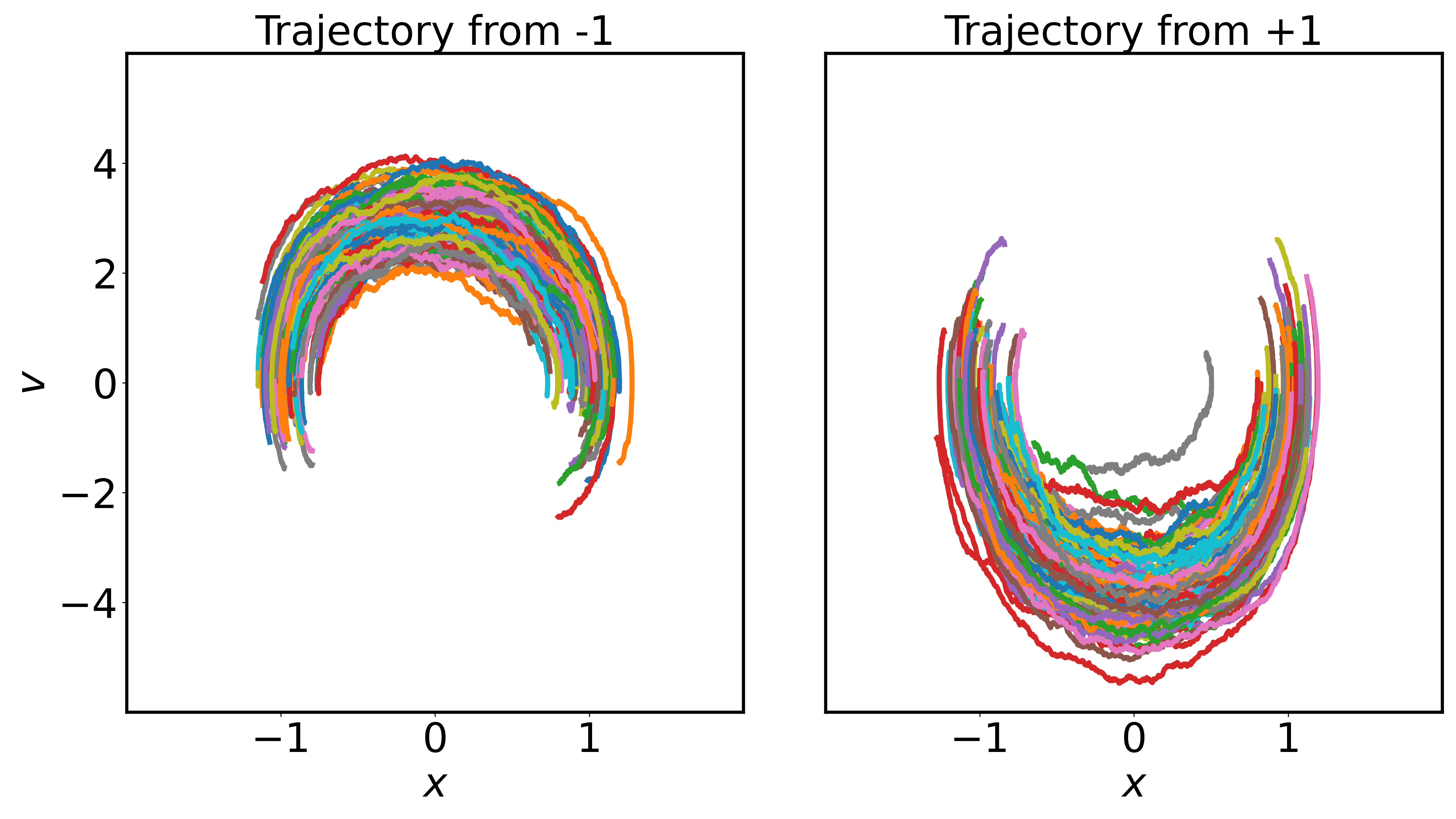}
    \end{subfigure}
\caption{(a) Bit-flip mean work: Initialization of the potential is $U(x,t)=0$ for $t\in(0,1)$. Gray dashed line is the work cost of protocol proposed in Ref. \cite{ray2021non}, which is $0.725 \kbt$. Trajectories are chosen from a single random training. (b) Phase space trajectories: Two trajectory classes in the whole phase space---one cluster starts near $-1$ and the other starts near $+1$.}
    \label{fig:flip_resultsI}
\end{figure}

\begin{figure}
    \centering
    \begin{subfigure}[t]{0.99\columnwidth}
    \subcaption{}
    \includegraphics[width=1\linewidth]
    {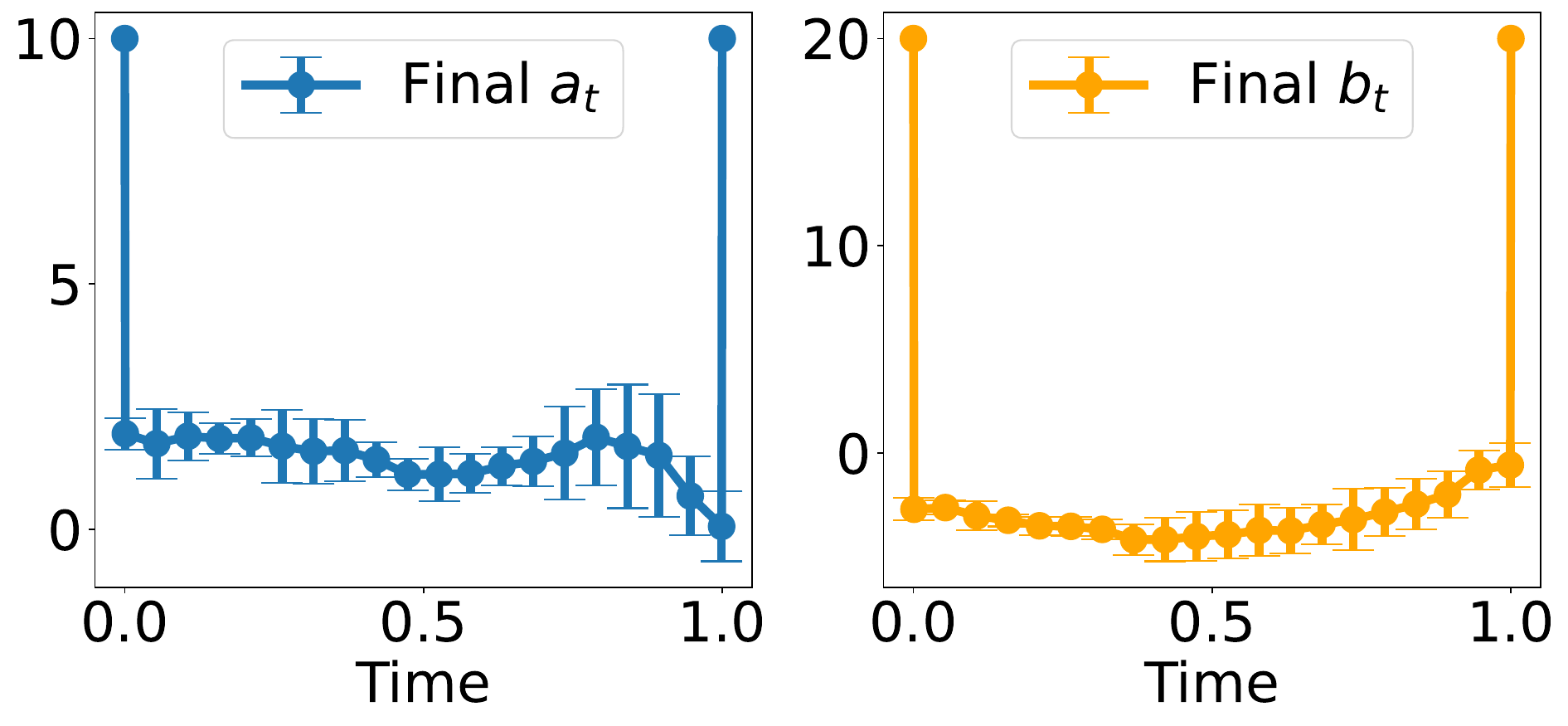}
    \end{subfigure}
    \begin{subfigure}[t]{0.99\columnwidth}
        \subcaption{}
        \includegraphics[width=1\linewidth]{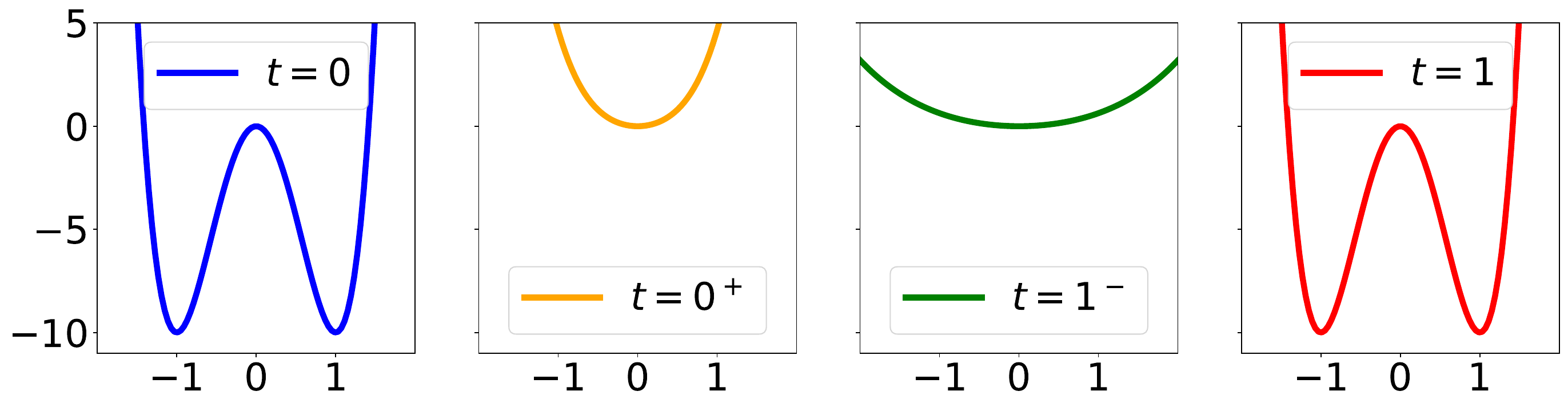}
    \end{subfigure}
\caption{(a) For each protocol value \{$a_{t}$, $b_t$\} in the double-well potential, we use $20$ break points for parameterization. They are initialized the same way as in bit erasure. After training on $1000$ iterations, we show the final protocol. (b) The potential jumps at $t=0^{+}$ and $t=1^{-}$.
%\jlnote{This caption makes no sense in what it is referring to: Top, middle, bottom panels?}
}
\label{fig:flip_resultsII}
\end{figure}

\paragraph*{Example: Underdamped-dynamics optimization and the bit-flip}

Let's turn to explore optimization in underdamped dynamics, a regime that is known to be challenging for existing methods. Position trajectories are no longer Markovian in underdamped dynamics and this property can be exploited to accomplish certain computational tasks that are unachievable in Markovian position dynamics (overdamped) without additional hidden states \cite{owen2019number}. An illustrative computational task is bit-flip in one-dimension. In this, trajectories originating from $-1$ and those from $+1$ must converge near the middle before reaching their respective final logical values. If the dynamics are purely Markovian, the computation loses the ability to distinguish between these two trajectory classes once they meet, making the bit-flip operation infeasible. In contrast, underdamped dynamics has additional degree of freedom through momentum, allowing one to distinguish histories and thus implement bit-flip operations effectively. 

Consider one-dimensional underdamped dynamics:
\begin{align*}
    dx = v dt ~ \text{and} ~ mdv = -\gamma v -  \partial_{x}U(x,a)dt+ \sigma dw_t~,
\end{align*}
where $\gamma$ is the damping coefficient and the thermal noise standard deviation is related to the thermal bath temperature $T$ by $\sigma = \sqrt{2\gamma k_{\text{B}}T}$. The probability of observing a trajectory $\omega=\{(x_t,v_t)\}_{t=0}^{\tau}$ with parameter $a$ is:
\begin{align*}
    \Pr(\omega,a)
    = \mathcal{N} \exp\left\{ - \frac{1}{2\sigma^2}\int_{\omega} \left[m \frac{dv}{dt} - \gamma v - \Phi(x,a) \right]^2 dt \right\}~,
\end{align*}
The FRR derived above in Eq. \eqref{eqn:FRR} follows directly from the chain rule and so the considerations above work just as well in underdamped regimes. Nothing in the optimization method changes under underdamped dynamics because $\partial_\alpha \log \Pr(\omega,a)$ takes exactly the same form here as in Eq.\eqref{eqn:FRR}.

The theoretical lower bound cost is $0 ~\kbt$ as the overall system entropy does not change during the bit-flip operation. Practically speaking, though, energy is dissipated in a finite-time protocol if there is nonzero damping. A more appropriate approximate lower bound is the energy dissipated during constant velocity transport across a distance $d$ from the bottom of the left well to the bottom of the right well: $W_{\min}\sim\gamma d^2 / \tau$.

To compare to our optimization, we use a heuristc bit-flip protocol introduced in Refs. \cite{ray2021non, ray2023gigahertz} and  physically implemented in Ref. \cite{dago2023logical}: By switching to an appropriately-chosen harmonic potential at $t=0^+$, the two clusters of particles exchange positions after one half-period of oscillation. We use the same potential, parameters $\boldsymbol{a}_{t}=\{a_{t}, b_{t}\}$, boundary conditions, and training initialization as in the bit-erasure task except the linear tilting term $c_t$ is removed. The error loss is chosen to be:
\begin{align}
\label{eqn:loss_ud_bitflip_mean}
    \mathcal{L}_{\text{flip}}(\boldsymbol{a}_t) =  \E{\omega}{\alpha_{1}\left(x_{\tau}-\text{target}(x_{0})\right)^2+\alpha_{W} W(\omega, \boldsymbol{a}_t)}~.
\end{align}
where $\text{target}(x_{0})=+1$ if $x_{0}<0$ and $\text{target}(x_{0})=-1$ if $x_{0}>0$. In numerical experiments, we set $\alpha_{1}=1$, work weight $\alpha_W=0.1$, $\gamma=0.1$, and $\tau=1$. Figures \ref{fig:flip_resultsI} and \ref{fig:flip_resultsII} present the training results. The work cost of our converged protocol is $\approx (0.615\pm0.070)\kbt$. This is close to the lower bound on the work cost based on constant velocity transport: $W_{\min}\sim0.4 \kbt$.

\section{Discussion and Conclusion}

Shifting from distributions to trajectories in contemporary nonequilibrium physics has led to many exciting results \cite{maes2020response, nicholson2020time, dieball2023direct, dieball2023feynman, dieball2024thermodynamic, melo2025stochastic, melo2025thermodynamic, dieball2025perspective, dieball2025precisely, dieball2025thermodynamic, fiusa2025counting, fiusa2025framework, barato2015thermodynamic,gingrich2016dissipation,horowitz2017proof, horowitz2020thermodynamic,liu2020thermodynamic,koyuk2020thermodynamic,lee2021universal,cao2022effective,potts2019thermodynamic,pietzonka2016universal, polettini2016tightening, proesmans2017discrete,dechant2018current,barato2018bounds,macieszczak2018unified,brandner2018thermodynamic,koyuk2019operationally, chun2019effect,van2019uncertaintydelayed,dechant2018multidimensional,hasegawa2019fluctuation,guarnieri2019thermodynamics,proesmans2019hysteretic,van2020uncertainty,timpanaro2019thermodynamic,hasegawa2020quantum,ray2023thermodynamic}. Building on this, we introduced a general framework for designing low-cost protocols for computational tasks embedded in physical systems. Given a computational task, we used the linear combination of work cost and statistical moments to construct the loss function and FRRs to compute the gradient and optimize the protocols. We demonstrated that the first and second-order statistical moments within the loss function effectively guide the system distribution toward the desired configuration in computational tasks.

More broadly, we used the Feynman–Kac formula to transform the optimization in partial differential equations to an optimization over trajectories. This reformulation offers a key advantage: simulating noisy trajectories is often much easier than solving the corresponding equations for the optimal protocol, either numerically or analytically. Our framework also shares similarity with reinforcement learning: a system (or agent) learns to optimize a protocol (or policy) by minimizing a loss function (or maximizing a reward signal) based on interactions with an environment \cite{sutton1998reinforcement}.

The error penalty used in Eq. \eqref{eqn: error_loss} is not unique. For example, the sign function $\text{sgn}(x)$ could also be used in the error penalty. Generally, this choice should depend on the specific requirements in the computational task. For example, if the final distribution $p_{\tau}$ should be exactly $p_{\text{target}}$, then a distribution distance as error penalty is needed; one could use the KL-divergence $\E{\omega\sim p_\tau}{\log \left(p_{\tau}/p_{\text{target}}\right)}$. However, computing the gradient of the $\log$ distribution $\partial_{a}\log p_{\tau}$ is typically challenging. A possible solution is to discretize the Fokker-Planck equation in time and iteratively solve for the distribution $p(x,t=i\Delta t;a)$ at each time step. Then one can use backpropagation to compute the gradient of KL-divergence with respect to $a$. Within the context of computational tasks, we observed that a limited number of statistical moments is effective for protocol design.

Notably, FRRs regained attention in recent years as studies found that they can bound the entropy production and dynamical activity in stochastic systems \cite{owen2020universal,chun2021nonequilibrium,aslyamov2024nonequilibrium,ptaszynski2024dissipation, zheng2024universal, zheng2025universal, gao2024thermodynamic, liu2025dynamical, aslyamov2025nonequilibrium,  bao2024nonequilibrium, van2025fundamental, kwon2025fluctuation, zheng2025nonlinear}. In line with this, our method reveals an application of FRRs---they are a tool to design low-cost protocols for probability-transform tasks within physical systems. The FRR-based gradient yields consistent results with back-propagation and genetic algorithms \cite{engel2023optimal, rengifo2025machine, whitelam2023train, casert2024learning, whitelam2025training, whitelam2025benchmarks, barros2025learning}. Perhaps equally important, the FRR method (i) offers a more-physical interpretation, (ii) requires significantly less memory storage, and (iii) is relatively easy to implement in code. Additionally, the generality of observables $J$ for which the FRR can provide a gradient means that the proposed framework can be extended well beyond information processing tasks simply by swapping out the loss function to encourage any behavior of choice.

The optimization framework presented here offers several promising future research directions and applications. First, it opens low-cost protocol design to different domains including unconventional computation \cite{aifer2024thermodynamic, pratt2025controlled, tang2025nonequilibrium, pratt2025extracting, zhang2025kinetic},  efficiency in information and nanoscale  engines \cite{boyd2016maxwell, boyd2017correlation}, chemical reactions \cite{chittari2023geometric}, biophyisical systems \cite{brown2019theory}, quantum gates \cite{van2024fidelity}, and the like. Second, having proved-out linear piecewise functions as our learning model, one future direction is to replace this with neural networks for enhanced scalability \cite{kim2020learning}. Third, we fixed the computation-time duration, but one can also treat the total computation time $\tau$ as a protocol parameter. Adding this to a loss function will enhance discovering protocols that balance duration along with error and work cost \cite{van2024time}.

\section*{Acknowledgments}

The authors thank Yibo Jacky Zhang for illuminating discussions on machine learning, as well as the Telluride Science and Innovation Center for its hospitality during visits and the participants of the Information Engines workshop there for their valuable feedback. This material is based on work supported by, or in part by, the U.S. Army Research Laboratory and U.S. Army Research Office under Grant No. W911NF-21-1-0048 and the Art and Science Laboratory.
\bibliography{ref}

\appendix

\section{Overdamped Harmonic Trap Translation}\label{appendix: translation_details}

This appendix gives the details needed to obtain the optimal protocol in overdamped harmonic trap translation. We denote the position mean as $\E{\omega}{x_{t}}=\mu_t$. First note that, by averaging the Langevin equation Eq. \eqref{eqn:move_sde}, we arrive at a simple constraint between $\mu_{t}$ and $a_{t}$:
\begin{align*}
\dot{\mu}_t + \mu_t =a_t 
~.
\end{align*}
From this, it is straightforward to see how the position mean at $t=\tau$ can be written in closed form:
\begin{align*}
    \mu_\tau \equiv \E{\omega}{x_\tau}&=\int_{0}^{\tau}e^{-(\tau-t)}a_t dt~,
\end{align*}
with the mean work being:
    \begin{align*}
    \E{\omega}{W}&=\int_{0}^{\tau} (a_t-\mu_t)\dot{a}_t dt~ . 
\end{align*}
The loss function is:
\begin{align*}
    \mathcal{L}_{\text{move}}(a_{t}) = \alpha_1\E{\omega}{(x_\tau-a_f)^2} + \alpha_{W}\E{\omega}{W(\omega,a_{t})}~.
\end{align*}
A helpful property of a linear force is that the variance of the distribution does not change with time $t$ if the distribution starts in equilibrium. So, we can safely replace $\E{\omega}{(x_\tau-a_f)^2}$ with $\left(\E{\omega}{x_\tau}-a_{f}\right)^2$ up to a constant. Adding a Lagrange multiplier $\lambda_{t}$ to enforce the constraint, and using the expressions for the means above, the function we want to minimize is:
\begin{align*}
    \mathcal{L}=&\alpha_{1}\left(\int_{0}^{\tau}e^{-(\tau-t)}a_t dt -a_f\right)^2 \\
    &+ \alpha_{w} \int_{0}^{\tau} (a_t-\mu_t)\dot{a}_t dt +\int_{0}^{\tau}\lambda_t(\dot{\mu}_t + \mu_t -a_t)dt~ .
\end{align*}
The Euler-Lagrange equations lead to three conditions:
\begin{align*}
    &a_{t}: 2\alpha_{1}e^{-(\tau-t)} \left(\int_{0}^{\tau}e^{-(\tau-s)}a_s ds -a_f\right) + \alpha_{w} \dot{\mu_{t}}-\lambda_{t}=0 ,\\
    &\mu_{t}:-\alpha_{w} \dot{a}_t + \lambda_{t} -\dot{\lambda}_{t} = 0 ~, ~\text{and}\\
    &\lambda_{t}: \dot{\mu}_t + \mu_t -a_t =0
    ~.
\end{align*}
We can eliminate the $\alpha_1$ term in the $a_{t}$ equation by solving for $\lambda_t$, and then subtracting the time derivative:
\begin{align*}
    \dot{\lambda}_{t} &=2\alpha_{1}e^{-(\tau-t)} \left(\int_{0}^{\tau}e^{-(\tau-s)}a_s ds -a_f\right) + \alpha_{w} \ddot{\mu}_{t}~.
\end{align*}
This yields the expression:
\begin{align}
\label{appendix_eq1}
    \lambda_{t} - \dot{\lambda}_t = \alpha_{w}\dot{\mu}_{t} - \alpha_{w}\ddot{\mu}_{t}~.
\end{align}
Taking the time derivative of the $\lambda_{t}$ equation and plugging it into the $\mu_{t}$ equation yields:
\begin{align}
    \lambda_{t} - \dot{\lambda}_t = \alpha_{w}\dot{\mu}_{t} + \alpha_{w}\ddot{\mu}_{t}~.
\end{align}
Thus, we find that $\ddot{\mu}_{t}=0$~. The initial condition $\mu_{0}=0$, then, implies $\mu_{t} = mt$ which reduces our constraint condition to $\lambda_{t}$ equation $a_{t} = mt +m$. The optimal constant $m^*$ depends on the hyperparameters, but can be determined exactly by minimizing the loss function $\mathcal{L}_{\text{move}}$ under the assumption of a linear $a_t$. The minimization yields:
\begin{align}
    m^{*}=a_{f}/(\tau+c),~\text{with}~ c =\frac{2\alpha_W}{2\alpha_1+\alpha_W}~.
\end{align}
Finally, we calculate the means of the final position and work when $m=m^{*}$:
\begin{align}
    &\E{\omega}{x_{\tau}}=\frac{a_{f}}{1+c/\tau}~,~\E{\omega}{W}={a _f^2}\frac{ \tau+c^2/2}{(\tau+c)^2}~.
\end{align}
We can see the effect of hyperparameters: increasing $\alpha_1$ (decreasing $c$) drives the system closer to the target value $a$. We recover a known result that takes only the work cost into account \cite{schmiedl2007optimal} when $\alpha_1\to0, \alpha_{W}\to 1$. As is often the case, the optimal protocol has jumps at the both ends.

\end{document}